\documentstyle[a4,11pt,psfig]{article}
\title{Ab-initio chemical potentials of solid and liquid solutions \\ 
and the chemistry of the Earth's core}

\author{D. Alf\`{e}$^{\star \,\dagger}$, M. J. Gillan$^\dagger$ and
G. D. Price$^\star$ \medskip \\ $^\star$Geological Sciences Department,
University College London \\ Gower Street, London WC1E 6BT, UK \medskip \\
$^\dagger$Physics and Astronomy Department, University College London
\\ Gower Street, London WC1E 6BT, UK}
\begin{document}
\date{ }
\maketitle

\begin{abstract}
A general set of methods is presented for calculating chemical
potentials in solid and liquid mixtures using {\em ab initio}
techniques based on density functional theory (DFT). The methods are
designed to give an {\em ab initio} approach to treating chemical
equilibrium between coexisting solid and liquid solutions, and
particularly the partitioning ratio of solutes between such
solutions. For the liquid phase, the methods are based on the general
technique of thermodynamic integration, applied to calculate the
change of free energy associated with the continuous interconversion
of solvent and solute atoms, the required thermal averages being
computed by DFT molecular dynamics simulation. For the solid phase,
free energies and hence chemical potentials are obtained using DFT
calculation of vibrational frequencies of systems containing
substitutional solute atoms, with anharmonic contributions calculated,
where needed, by thermodynamic integration. The practical use of the
methods is illustrated by applying them to study chemical equilibrium
between the outer liquid and inner solid parts of the Earth's core,
modelled as solutions of S, Si and O in Fe. The calculations place
strong constraints on the chemical composition of the core, and allow
an estimate of the temperature at the inner-core/outer-core boundary.
\end{abstract}

\section{Introduction}
\label{sec:intro}
We present here a set of techniques that allow the {\em ab initio}
calculation of chemical potentials in solid and liquid solutions,
and hence the {\em ab initio} treatment of chemical equilibrium
between solid and liquid phases. There are many areas of chemical
physics where such techniques might be important, but we believe
they have a particular role to play in studying the partitioning of
impurities between different phases under extreme conditions, where
experiments are difficult or impossible. As an illustration of the
power of the techniques, we will describe how we have applied them
to study chemical equilibrium between the solid and liquid parts of
the Earth's core. 

The techniques to be presented form a natural
sequel to recent developments in the {\em ab initio} thermodynamics
of condensed matter based on electronic density-functional theory
(DFT)~\cite{generaldft}. For many years, DFT has been used to
calculate the phonon spectra of perfect crystals~\cite{baroni87},
and it is only a short step from that to the calculation of free
energies and other thermodynamic quantities in the harmonic
approximation. There have already been several reports of DFT
calculations of high-temperature crystal
thermodynamics~\cite{karki00,Lichtenstein00,xie99a,xie99b,
lazzeri98,christensen00a,alfe01a}, including solid-solid phase
equilibria~\cite{pavone98,kern99,gaal-nagy99,christensen00b} using
this approach. For liquids, {\em ab initio} thermodynamics first
became possible with the Car-Parrinello technique~\cite{car85} of
DFT molecular dynamics simulation, which immediately gave a way to
calculate such quantities as pressure, internal energy and
temperature of a liquid in thermal equlibrium.  The first DFT
treatment of solid-liquid equilibrium was achieved by Sugino and
Car~\cite{sugino95}, who used thermodynamic integration to compute
the free energies of solid and liquid Si, and hence the melting
properties of the material. Closely related is the work of de~Wijs
{\em et al.}~\cite{dewijs98} on the melting of Al. We have recently
reported DFT calculations of the free energies and melting curves of
Fe~\cite{alfe01a,alfe99a,alfe01c} and Al~\cite{vocadlo01} over a
wide range of pressures; Jesson and Madden~\cite{jesson2000} have
recently presented {\em ab initio}  calculations of the
zero-pressure melting properties of Al using their `orbital free'
approach. The work of Smargiassi and Car~\cite{smargiassi96} and
Smargiassi and Madden~\cite{smargiassi95} on the free energy of
formation of defects in crystals is also relevant to the ideas to be
presented here. 

Thermodynamic integration has been the key to
calculating the {\em ab initio} free energy of liquids and
anharmonic solids, and hence to the treatment of solid-liquid
equilibrium.  It provides a means of computing the difference of
free energy between the {\em ab initio} system and a reference model
system whose free energy is known. We will show that it is also the
key to calculating {\em ab initio} chemical potentials of liquids
and anharmonic solids, but here it is used in a rather different
way. The chemical potential of a species is the free energy change
when an atom of that species is added to the system.  The difference
of chemical potentials of two species is therefore the free energy
change when an atom of one species is replaced by an atom of the
other, or equivalently when one atom is transmuted into the
other. The role of thermodynamic integration here is to provide a
way of calculating the free energy change associated with such
transmutations, and we shall show how this can be accomplished in
practical DFT simulations. This general approach is closely related
to ideas that have been used for a long time in classical
simulation. A recent example of classical thermodynamic integration
with molecular transmutation to calculate solvation free energies in
aqueous solution can be found in Ref.~\cite{arthur98}, which gives
references to earlier literature. 

Although the techniques we
shall present are fairly general, we do impose two restrictions at
present: First, the theoretical framework is developed for the case
of a two-component mixture; second, one of the components is present
at low, but not very low, concentration, in a sense to be clarified
in Sec.~\ref{sec:chemequ}. The situation envisaged therefore
consists of fairly dilute solid and liquid solutions in
coexistence. 

There are vast numbers of problems both in the
chemical industry and in the natural world that depend on the
partitioning of chemical components between coexisting phases, and
the ability to calculate chemical potentials {\em ab initio} should
make it possible to address some of these problems in a new way. Our
original incentive for developing these techniques was the desire to
understand better the chemistry of the Earth's core, and this is a
good example of a problem where {\em ab initio} calculations can
supply information that is difficult to obtain experimentally
because of the extreme conditions of temperature ($T \sim 6000$~K)
and pressure ($p \sim 330$~GPa). The core is composed mainly of Fe,
and comprises an outer liquid part and an inner solid
part~\cite{birch64}.  The density of the outer core is $\sim 6$~\%
too low to be pure
Fe~\cite{birch64,ringwood77,poirier94,alfe00b,laio00}, and
cosmochemical and geochemical arguments show that the main light
impurities are probably S, O and Si~\cite{poirier94}. The inner core
has grown over geological time by crystallisation from the outer core,
and the partitioning of impurities between liquid and solid is
crucial for understanding the evolution and contemporary dynamics of
the core. The size of the density discontinuity ({\em ca.}
4.5~\%)~\cite{shearer90,prem} at the inner-core/outer-core boundary
can only be interpreted if one understands this partitioning, and
also provides a constraint on possible chemical compositions. We
shall show how our {\em ab initio} techniques for calculating
chemical potentials shed completely new light on this problem. Brief
reports of these calculations have already
appeared~\cite{alfe00b,alfe00c,alfe01b}. 

In developing the
theoretical basis of our techniques, we define our technical aims in
Sec.~\ref{sec:chemequ} by summarising the standard thermodynamic
relations describing phase equilibrium. The difference of chemical
potentials of solute and solvent atoms, and the free energy change
associated with the transmutation of solvent into solute are
discussed in Sec.~\ref{sec:interconversion}. In
Sec.~\ref{sec:chempot}, we then develop the {\em ab initio}
techniques themselves. We shall explain how thermodynamic
integration can be used to perform the solvent-solute transmutation
so as to obtain the difference of  chemical potentials in the
liquid; we also  describe the techniques for calculating chemical
potentials in solid solutions, both in the harmonic approximation
and using thermodynamic integration to handle
anharmonicity. Sec.~\ref{sec:earths_core} presents our results for
the case of S, O and Si dissolved in solid and liquid Fe under
Earth's core conditions, and summarizes the implications of the
results for the partitioning of these impurities between the inner
and outer core and the chemical composition of the core. Discussion
and conclusions are given in Sec.~\ref{sec:discuss}.

\section{Chemical equilibrium: thermodynamics}
\label{sec:chemequ}

Our task in this Section is to identify the thermodynamic
quantities that will need to be calculated {\em ab initio}.
Chemical equilibrium between two multicomponent phases is characterized
by equality of the chemical potentials of each component in the
two phases. For a two-component solution consisting of solute~X
dissolved in solvent~A, equilibrium between solid and liquid phases
requires that:
\begin{eqnarray}
\mu_{\rm X}^l(p,T_m,c^l_{\rm X}) = \mu_{\rm X}^s(p,T_m,c^s_{\rm X}) \; , 
\label{eqn:muxmux} \\
\mu_{\rm A}^l(p,T_m,c^l_{\rm X}) = \mu_{\rm A}^s(p,T_m,c^s_{\rm X}) \; ,
\label{eqn:mufemufe}
\end{eqnarray}
where $\mu_{\rm X}$ and $\mu_{\rm A}$ are the chemical potentials
of solute and solvent,
$p$ is the pressure and $c_{\rm X}$ is the mole fraction of solute,
with superscripts $l$ and $s$ for liquid and solid
respectively; $T_m$ is the melting temperature, i.e. the temperature
at which the liquid and solid solutions are in equilibrium, which
depends on the impurity mole fractions.
The two equations impose two relations
between $c_{\rm X}^l$, $c_{\rm X}^s$ and $T_m$ at the given $p$.
In the low-concentration limit $c_{\rm X} \rightarrow 0$,
$\mu_{\rm X}$ diverges logarithmically, and it is useful to write:
\begin{equation}\label{eqn:chempotsolute}
\mu_{\rm X}(p,T,c_{\rm X}) = k_{\rm B} T \ln c_{\rm X} +
\tilde{\mu}_{\rm X}(p,T,c_{\rm X}) \; ,
\end{equation}
where $\tilde{\mu}_{\rm X} (p,T,c_{\rm X})$ is well behaved 
for all $c_{\rm X}$. In an ideal solution,
$\tilde{\mu}_{\rm X}$ is independent of $c_{\rm X}$, but in
reality the interaction between solute atoms causes it to vary
with $c_{\rm X}$. Combining
Eqs.~(\ref{eqn:muxmux}) (\ref{eqn:chempotsolute}), we obtain:
\begin{equation}\label{eqn:partition}
c_{\rm X}^s / c_{\rm X}^l = \exp [ ( {\tilde{\mu}}_{\rm X}^l -
{\tilde{\mu}}_{\rm X}^s ) / k_{\rm B} T_m ] \; ,
\end{equation}
so that the ratio of the mole fractions $c_{\rm X}^s$ and
$c_{\rm X}^l$ in solid and liquid is determined by the
thermodynamic quantities ${\tilde{\mu}}_{\rm X}^l$ and
${\tilde{\mu}}_{\rm X}^s.$ 
The melting temperature $T_m$ entering
Eq.~(\ref{eqn:partition}) differs from the melting 
temperature $T_m^0$ of the pure
solvent, and may be regarded as determined by Eq.~(\ref{eqn:mufemufe}).

We now develop a practical way of solving Eqs.~(\ref{eqn:mufemufe})
and (\ref{eqn:partition}).
We are interested in the case of moderately low $c_{\rm X}$, but
we wish to take account of the variation of ${\tilde{\mu}}_{\rm X}$
with $c_{\rm X}$ to lowest order. We therefore expand
${\tilde{\mu}}_{\rm X}$ as:
\begin{equation}\label{eqn:mux}
{\tilde{\mu}}_{\rm X}(p,T,c_{\rm X}) = \mu^\dagger_{\rm X}(p,T) + 
\lambda_X(p,T) c_{\rm X} +  O(c_{\rm X}^2) \; ,
\end{equation}
and we shall systematically neglect the terms $O ( c_{\rm X}^2 )$.
Since it will be important later, we stress here that this
represents the concentration dependence of ${\tilde{\mu}}_{\rm X}$
at constant pressure.
Eq.~(\ref{eqn:partition}) then becomes: 
\begin{equation}\label{eqn:partition_2}
c_{\rm X}^s / c_{\rm X}^l = \exp \left[ ( \mu^{\dagger l}_{\rm X} -
\mu^{\dagger s}_{\rm X} + \lambda_X^l c^l_{\rm X} - \lambda_X^s c^s_{\rm
X} ) / k_{\rm B} T_m \right] \; .
\end{equation}
To obtain an equation for $T_m$, we need the
corresponding expansion for $\mu_{\rm A}$. We use the 
Gibbs-Duhem equation:
\begin{equation}
c_{\rm A} {\rm d}\mu_{\rm A} + c_{\rm X} {\rm d}\mu_{\rm X} = 0 \; ,
\end{equation}
which gives:
\begin{eqnarray}
{\mu}_{\rm A}(p,T,c_{\rm X}) = \mu^0_{\rm A}(p,T) + 
(k_{\rm B} T + \lambda_{\rm X}(p,T)) \ln (1-c_{\rm X}) + 
\lambda_{\rm X}(p,T) c_{\rm X} +
O(c_{\rm X}^2) \; ,
\end{eqnarray}
where $\mu_{\rm A}^0$ is the chemical potential of the
pure solvent, and
we have used the fact that $c_{\rm A} = 1 - c_{\rm X}$. To linear
order in $c_{\rm X}$, this gives:
\begin{equation}\label{eqn:mufe}
{\mu}_{\rm A}(p,T,c_{\rm X}) =  
\mu^0_{\rm A}(p,T) - k_{\rm B} T c_{\rm X} +  O (c_{\rm X}^2) \; .
\end{equation}
We apply this in Eq.~(\ref{eqn:mufemufe}) by expanding
$\mu_{\rm A}^0 ( p , T )$ to linear order in
the difference $T_m - T_m^0$ between $T_m$ and the melting
temperature $T_m^0$ of pure solvent. This yields:
\begin{eqnarray}
- k_{\rm B} T_m c_{\rm X}^l +
\mu^{0 l}_{\rm A}(p,T^0_m) + (T_m - T^0_m) 
\left( \frac{\partial \mu^{0 l}_{\rm A}}{\partial T} \right )_{T=T^0_m}
& = & \nonumber \\
- k_{\rm B} T_m c_{\rm X}^s +
\mu^{0 s}_{\rm A}(p,T^0_m) & + & (T_m - T^0_m) 
\left ( \frac{\partial \mu^{0 s}_{\rm A}}{\partial T} \right )_{T=T^0_m}
\; .
\end{eqnarray}
Since $\mu^{0 s}_{\rm A}(p,T^0_m) = 
\mu^{0 s}_{\rm A}(p,T^0_m)$, we can rewrite this equation as:
\begin{equation}
k_{\rm B} T_m c_{\rm X}^l + (T_m - T^0_m) s_{\rm A}^{0 l} = 
k_{\rm B} T_m c_{\rm X}^s + (T_m - T^0_m) s_{\rm A}^{0 s}, 
\end{equation}
where $s_{\rm A}^0 = -\left( \partial \mu^0_{\rm A} / 
\partial T \right)_{T=T^0_m}$ is the entropy 
per atom of pure solvent at
the melting temperature. The shift of melting temperature due to
the presence of the solute is then:
\begin{equation}\label{eqn:Tm_shift}
(T_m - T^0_m) = 
\frac{k_{\rm B} T_m}{ \Delta s_{\rm A}^0} ( c_{\rm X}^s  - c_{\rm X}^l ) \; ,
\end{equation}
where $\Delta s_{\rm A}^0 \equiv s_{\rm A}^{0 l} - s_{\rm A}^{0 s}$ 
is the entropy of fusion of
pure solvent. Eqs.~(\ref{eqn:partition_2}) and~(\ref{eqn:Tm_shift}) must
be solved self-consistently.

We see from Eqs.~(\ref{eqn:partition_2}) and (\ref{eqn:Tm_shift})
that the main thermodynamic quantities to be calculated
{\em ab initio} are $\mu_{\rm X}^\dagger$ and $\lambda_{\rm X}$
for the solid and liquid solution. We also require
{\em ab initio} values for the melting temperature and entropy
of fusion of pure solvent. In addition, we shall find it necessary
to obtain {\em ab initio} values of the partial molar volumes
of the solute, which will be discussed later. 
We next turn to the statistical-mechanical
considerations needed to develop a strategy for calculating 
$\mu_{\rm X}^\dagger$ and $\lambda_{\rm X}$.

\section{Interconversion of solvent and solute: statistical mechanics}
\label{sec:interconversion}

The chemical potential of a solute X in solid or liquid solvent A is
the change of Gibbs free energy when one atom of X is added to
the system at constant pressure and temperature.
In our practical {\em ab initio} calculations, we work
at constant volume rather than constant pressure, so we
prefer the equivalent statement that the solute chemical
potential is the change of Helmholtz free
energy when the solute atom is added at constant volume and
temperature. However, it is impractical to add solute atoms to a system
in an {\em ab-initio} simulation.  It is more
convenient to convert solvent into solute, which means working with the
difference $\mu_{\rm X} - \mu_{\rm A}$ of the chemical potentials.
The chemical potential $\mu_{\rm X}$ and hence ${\tilde{\mu}}_{\rm
X}$ can then be obtained from $\mu_{\rm X} - \mu_{\rm A}$ by making use of the
Gibbs free energies of pure solid and liquid A, calculated separately.
We have sketched this technique briefly in our previous
papers~\cite{alfe00b,alfe00c,alfe01b}, and we now describe it 
in more detail.

The Helmholtz free energy of a system containing $N_{\rm A}$ solvent
atoms and $N_{\rm X}$ solute atoms is:
\begin{equation}
F(N_{\rm A}, N_{\rm X}) = -k_{\rm B}T \ln \left \{ \frac{1}{N_{\rm
A}!N_{\rm X}! \Lambda_{\rm A}^{3N_{\rm A}} \Lambda_{\rm
X}^{3N_{\rm X}}} \int_V {\rm d}{\bf R} \exp\left [ -\beta U(N_{\rm A},
N_{\rm X};{\bf R} ) \right ] \right \} \; ,
\label{eqn:helmholtz}
\end{equation}
where $\beta = 1/k_{\rm B}T$, and $\Lambda_{\rm X}$ and
$\Lambda_{\rm X}$ are the thermal wavelengths of A and X, given
by $\Lambda_{\rm A} = h / ( 2 \pi M_{\rm A} k_{\rm B} T )^{1/2}$,
with $M_{\rm A}$ the atomic mass of A, and similarly for
$\Lambda_{\rm X}$. The quantity $U(N_{\rm A}, N_{\rm X};{\bf R} )$ is the 
total energy function of the system of $N_{\rm A}$ solvent
and $N_{\rm X}$ solute atoms, which
depends on the positions of all the atoms, indicated by {\bf R}, and
$\int_V  {\rm   d}{\bf  R}$  indicates integration over the whole
configuration space of the system contained in volume $V$. 

The difference of chemical potentials $\mu_{XA} \equiv \mu_{X} - \mu_{A}$
is equal to the change of $F$ when a single atom of A is 
converted into X, and is given by:
\begin{eqnarray}
\mu_{XA} & = & F(N_{\rm A} - 1 , N_{\rm X} +1) - F(N_{\rm A}, N_{\rm X})
\nonumber \\ 
 & = & -k_{\rm B}T \ln ( N_{\rm A} / N_{\rm X} ) 
    -k_{\rm B}T \ln \left( \Lambda_{\rm A}^3 / 
\Lambda_{\rm X}^3 \right) \nonumber \\
 & \; & \; \; \; \; -  k_{\rm B} T \ln \left\{
\frac{\int_V {\rm d}{\bf R} \exp\left[ -\beta U(N_{\rm A}-1, N_{\rm
X}+1;{\bf R} ) \right] }{\int_V {\rm d}{\bf R} \exp\left[ -\beta
U(N_{\rm A}, N_{\rm X};{\bf R} ) \right] }\right\} \; .
\label{eqn:mu_xa1}
\end{eqnarray}
We express this as:
\begin{eqnarray}
\mu_{XA} = k_{\rm B}T \ln\frac{c_{\rm X}}{1 - c_{\rm X}} + 3 k_{\rm
B}T \ln \frac{\Lambda_{\rm X}}{\Lambda_{\rm A}} + m ( c_{\rm X} ) \; ,
\label{eqn:mu_xa2}
\end{eqnarray}
where we define:
\begin{equation}\label{eqn:m}
m(c_{\rm X}) = -k_{\rm B}T \ln \left \{ \frac{\int_V {\rm d}{\bf R}
\exp\left [ -\beta U(N_{\rm A}-1, N_{\rm X}+1;{\bf R} ) \right ]
}{\int_V {\rm d}{\bf R} \exp\left [ -\beta U(N_{\rm A}, N_{\rm X};{\bf
R} ) \right ] }\right \} \; .
\end{equation}
The intensive quantity $m ( c_{\rm X} )$ 
depends only on pressure, temperature and concentration (we
write it as $m ( p , T , c_{\rm X} )$), or alternatively on
volume, temperature and concentration (we then write it
as $m ( \bar{v} , T , c_{\rm X} )$, where $\bar{v}$ is
the mean atomic volume $V / (N_{\rm A} + N_{\rm X} ) $).
Expanding Eq.~(\ref{eqn:mu_xa2}) to linear order in $c_{\rm X}$, we have:
\begin{equation}
\mu_{XA} = k_{\rm B}T \ln c_{\rm X} + k_{\rm B}T c_{\rm X} + 3 k_{\rm
B}T \ln \frac{\Lambda_{\rm X}}{\Lambda_{\rm A}} + m(c_{\rm X}).
\end{equation}
Compare now with Eqs.~(\ref{eqn:chempotsolute}), 
(\ref{eqn:mux}) and (\ref{eqn:mufe}):
\begin{equation}
\mu_{XA} = k_{\rm B} T \ln c_{\rm X} + \mu^\dagger_{\rm
X} + \lambda_X c_{\rm X} - \mu^0_{\rm A} + k_{\rm B} T c_{\rm X} \; ,  
\end{equation}
and we have:
\begin{equation}\label{eqn:m_2}
m(c_{\rm X}) + 3 k_{\rm B}T \ln \frac{\Lambda_{\rm X}}
{\Lambda_{\rm A}} = \mu^\dagger_{\rm X}  - \mu^0_{\rm A} + 
\lambda_X c_{\rm X}.
\end{equation}
If values are available for 
$m ( p , T , c_{\rm X} )$ at different values of $c_{\rm X}$
for a given pressure $p$, we can obtain the quantities
$\mu_{\rm X A}^\dagger \equiv \mu_{\rm X}^\dagger - \mu_{\rm A}^0
- 3 k_{\rm B} T \ln ( \Lambda_{\rm X} / \Lambda_{\rm A} )$
and $\lambda_{\rm X}$ for that pressure. We then need the pure-solvent
chemical potentials $\mu_{\rm A}^0$ for liquid and solid,
whose {\em ab initio} calculation has been described in 
detail elsewhere~\cite{alfe01a,alfe99a,alfe01c}.

We conclude this Section by rewriting $c_{\rm X}^s / c_{\rm X}^l$ from
Eq.~(\ref{eqn:partition}) in terms of 
$\mu_{\rm X A}^\dagger$, $\lambda_{\rm X}$
and $\mu_{\rm A}^0$ for liquid and solid:
\begin{equation}
c_{\rm X}^s / c_{\rm X}^l = \frac{ \exp \left[ \left(
\mu_{\rm X A}^{\dagger l} - \mu_{\rm X A}^{\dagger s} +
\lambda_{\rm X}^l c_{\rm X}^l - \lambda_{\rm X}^s c_{\rm X}^s \right) /
k_{\rm B} T_m \right] }
{ \exp \left[ \left( 
\mu_{\rm A}^{0 s} - \mu_{\rm A}^{0 l} \right) /
k_{\rm B} T_m \right] } \; .
\end{equation}
If the concentrations are small enough for the difference
between $T_m$ and $T_m^0$ to be negligible, then
$\mu_{\rm A}^{0 s} = \mu_{\rm A}^{0 l}$, and the denominator is
unity; but in general deviations of the denominator from
unity should be included.

%

\section{{Ab initio} chemical potentials}
\label{sec:chempot}

\subsection{The liquid solution}
\label{sec:liqsol}

For the liquid, we calculate the quantity $m ( c_{\rm X} )$ of
Eq.~(\ref{eqn:m}) by a form of `thermodynamic integration'. We first
outline a simple way of doing this that is correct in
principle, but suffers from practical problems; we then show
how the method can be modified to give a practical procedure.

Thermodynamic integration~\cite{frenkel96} is a general
technique for computing the Helmholtz free energy difference $F_1 -
F_0$ of two systems containing the same number $N$ of atoms, but
having different total energy functions $U_1 ( {\bf R} )$
and $U_0 ( {\bf R} )$. The difference $F_1 - F_0$ is the
reversible work done on continuously switching the total energy
function from $U_0$ to $U_1$ at constant volume, which is given by
\begin{equation}\label{eqn:TI}
F_1 - F_0 = \int_0^1 d \lambda \, \langle U_1 - U_0 \rangle_\lambda \; ,
\end{equation}
where the average $\langle \, \cdot \, \rangle$ is calculated in
thermal equilibrium for the system governed by the `hybrid' energy
function $U_\lambda \equiv ( 1 - \lambda ) U_0 + \lambda U_1$. This is
a well established technique for the {\em ab initio} calculation of
liquid-state free energies~\cite{sugino95,dewijs98}, which was used in
our recent {\em ab initio}
investigation~\cite{alfe01a,alfe99a,alfe01c} of the high pressure
melting curve of Fe.  

In order to compute $m ( c_{\rm X} )$, we could in principle 
choose $U_0$ to be the total energy for the 
system of $N_{\rm A}$ atoms of solvent and
$N_{\rm X}$ of solute, and $U_1$ to be the same for $N_{\rm A} -1$
atoms of A and $N_X + 1$ of X. We evaluate $\langle U_1 - U_0
\rangle_\lambda$ by performing {\em ab initio} molecular dynamics with
time evolution generated by $U_\lambda$, and taking the time average
of $U_1 - U_0$. This is repeated for several values of $\lambda$, and
the integration over $\lambda$ is done numerically.
This type of `alchemical' transmutation of A into X obviously does not
correspond to a real-world process, but in terms of {\em ab initio}
statistical mechanics is a perfectly rigorous way of obtaining the
quantity $m ( c_{\rm X} )$.  It demands an 
unusual kind of simulation: For the atom
positions ${\bf r}_1 , \ldots {\bf r}_N$ at each instant of time, we
have to perform two independent {\em ab initio} calculations, one for
each chemical composition. As well as $U_0$ and $U_1$ for the given
positions, we calculate two sets of {\em ab initio} forces ${\bf F}_{0
i} \equiv - \nabla_i U_0$ and ${\bf F}_{1 i} \equiv - \nabla_i U_1$,
and the linear combinations ${\bf F}_{\lambda i} \equiv ( 1 - \lambda
) {\bf F}_{0 i} + \lambda {\bf F}_{1 i}$ are used to generate the time
evolution.  

The major problem with this scheme is one of statistics. The thermal
average $\langle \, \cdot \, \rangle_\lambda$ 
is evaluated as a time average,
but since only a single atom is transmuted the scheme does not
benefit from averaging over atoms. The efficiency of the averaging
can be considerably improved if one is prepared to transmute 
several atoms simultaneously. If we do this, then instead of
obtaining $m ( c_{\rm X} )$ at a given
mole fraction $c_{\rm X}$, we obtain an integral of $m ( c_{\rm X} )$
over a range of $c_{\rm X}$ values. The information we need
can still be extracted, as we now describe.

Consider the change of Helmholtz free energy when we start from
$N$ atoms of pure solvent and transmute $N_{\rm X}$ of them
into solute atoms at constant volume and temperature. 
This can clearly be calculated by thermodynamic integration
using the procedure outlined above. Denoting this
change of free energy by $W ( N , N_{\rm X} )$, we can express it as:
\begin{equation}
W(N,N_{\rm X}) = -k_{\rm B}T \ln \left\{ \frac{\int_V {\rm d}{\bf R}
\exp\left[ -\beta U(N-N_{\rm X},N_{\rm X};{\bf R} ) \right]
}{\int_V {\rm d}{\bf R} \exp\left[ -\beta U(N, 0;{\bf
R} ) \right] }\right\}.
\end{equation}
We then have:
\begin{equation}
W ( N , N_{\rm X} ) = \int_0^1 d \lambda \,
\langle U_1 - U_0 \rangle_\lambda \; ,
\label{eqn:thermint_W}
\end{equation}
with $U_1 ( {\bf R} ) = U ( N - N_{\rm X} , N_{\rm X} ; {\bf R} )$ and
$U_0 ( {\bf R} ) = U ( N , 0 ; {\bf R} )$.
Our procedure will be to calculate $W ( N , N_{\rm X} )$
at several values of $N_{\rm X} / N = c_{\rm X}$
at a chosen volume, and then fit the results in the following
way:
\begin{equation}\label{eqn:w}
W ( N , N_{\rm X} ) / N_{\rm X} = a + b c_{\rm X} \; .
\end{equation}
The information needed can now be extracted by noting that
for a given mean atomic volume $\bar{v}$, the quantity
$m ( \bar{v} , T , c_{\rm X} )$ is:
\begin{equation}
m ( \bar{v} , T , c_{\rm X} ) =
( \partial W / \partial N_{\rm X} )_{V , T} =
a + 2 b c_{\rm X} \; .
\end{equation}
It follows immediately that:
\begin{equation}
\mu_{\rm X A}^\dagger = \lim_{c_{\rm X} \rightarrow 0}
m ( \bar{v} , T , c_{\rm X} ) = a \; .
\end{equation}
To obtain $\lambda_{\rm X}$ from the coefficient $b$, we
note that $\lambda_{\rm X} = \lim_{c_{\rm X} \rightarrow 0}
( \partial m ( p , T , c_{\rm X} ) / \partial c_{\rm X} )_p$ and
$2 b = \lim_{c_{\rm X} \rightarrow 0}
( \partial m ( \bar{v} , T , c_{\rm X} / 
\partial c_{\rm X} )_{\bar{v}}$.
The fact that one derivative is isobaric and the other isochoric
is significant. The quantity $\lambda_{\rm X}$ that we seek
describes the isobaric concentration dependence of solute chemical
potential. But since our {\em ab initio} calculations 
are done at fixed volume, the immediately available quantity $b$ 
is an isochoric derivative.

The relation between the constant-pressure and constant-volume
derivatives of $m$ is examined in the Appendix, where we show that:
\begin{equation}
( \partial m / \partial c_{\rm X} )_p =
( \partial m / \partial c_{\rm X} )_{\bar{v}} -
n B_T ( v_{\rm X} - v_{\rm A} )^2 \; ,
\end{equation}
where $B_T$ is the isothermal bulk modulus, and $v_{\rm X}$
and $v_{\rm A}$ are the partial atomic volumes of solute and
solvent. We conclude that:
\begin{equation}\label{eqn:lambda_correction}
\lambda_{\rm X} = 2 b - n B_T ( v_{\rm X} - v_{\rm A} )^2 \; .
\end{equation}
Here, the quantities $B_T$, $v_{\rm X}$ and $v_{\rm A}$ can be evaluated
at infinite dilution. The calculation of $B_T$ and $v_{\rm A}$
involves only {\em ab initio} m.d. simulations on the pure
solvent, and presents no problem.
We return below (Sec.~\ref{sec:partial_molar_volume}) 
to the {\em ab initio} calculation of $v_{\rm X}$. With this,
we have a complete procedure for determining 
$\mu_{\rm X A}^\dagger$ and $\lambda_{\rm X}$.

\subsection{The solid solution}
\label{sec:solsol}

If anharmonic effects are negligible, then the free energy
of the solid can be obtained from {\em ab initio} phonon
frequencies, so that thermodynamic integration is not needed,
and no statistical averaging is involved in the {\em ab initio}
calculations. There is then nothing to prevent us from calculating
$m ( c_{\rm X} )$ directly from the free energy change when 
solvent atoms are replaced by solute atoms. We assume for the moment
that this is adequate, and return later to the question of
anharmonic effects.

We start by considering the zero-concentration limit of
$m ( c_{\rm X} )$, namely $\mu_{\rm XA}^\dagger$, which is
the non-configurational free energy change when an atom in
the perfect A crystal is replaced by an X atom. This can be
written as:
\begin{equation}
\mu_{\rm XA}^\dagger = \mu_{\rm XA}^{\dagger \, {\rm perf}} +
\mu_{\rm XA}^{\dagger \, {\rm harm}} \; ,
\end{equation}
where $\mu_{\rm XA}^{\dagger \, {\rm perf}}$ is the free energy
change for the perfect non-vibrating crystal, and
$\mu_{\rm XA}^{\dagger \, {\rm harm}}$ is the harmonic vibrational
part -- we refer to $\mu_{\rm XA}^{\dagger \, {\rm perf}}$ as
a `free energy' to allow for the possibility of thermal
electronic excitations, which are important in 
high-temperature Fe~\cite{alfe01a}.
The calculation of $\mu_{\rm XA}^{\dagger \, {\rm perf}}$ is straightforward,
and involves only the difference of {\em ab initio} (free) energies
of the static fully relaxed crystal containing a single substitutional
X atom and the static perfect crystal, the two systems having
the same volume.

In the high-temperature limit, where $T$ is well above the Debye
temperature, $\mu_{\rm XA}^{\dagger \, {\rm harm}}$
can be written as:
\begin{equation}
\mu_{\rm XA}^{\dagger \, {\rm harm}} = k_{\rm B} T
\sum_n \ln \left( \omega_n^\prime / \omega_n \right) \; ,
\label{eqn:mu_harm}
\end{equation}
where $\omega_n^\prime$ and $\omega_n$ are the harmonic frequencies
of the normal modes of the impure and pure crystals, and the sum
goes over all modes. The frequencies are calculated {\em ab initio},
and we use the `small-displacement' method described in detail
elsewhere~\cite{alfe01a,kresse95,darioweb}. This involves DFT calculations of
the force on every atom in the system induced by displacement of a
single atom, and this has to be done for all symmetry inequivalent
atoms and displacements. To obtain
$\lambda_{\rm X}$ in the harmonic approximation, we must include the
effect of interactions between solute atoms. The key to this is to
note that the calculation of the partition function, i.e. the
integral over configuration space of Eq.~(\ref{eqn:helmholtz}), can
be broken into (a) a sum of distinct configurations,
i.e. assignments of solute atoms to lattice sites, and (b) an
integral over vibrational displacements of the atoms away from their
relaxed equilibrium positions for each such configuration. This
means that the statistical mechanics of the solid solution maps
exactly onto that of a lattice gas, and the free energy of the solid
solution is:
\begin{equation}
F ( N_{\rm A} , N_{\rm X} ) = - k_{\rm B} T
\ln \left( \sum_\gamma e^{- \beta \Phi_\gamma} \right) \; ,
\end{equation}
where the sum goes over all distinct configurations $\gamma$,
and $\Phi_\gamma$ is the non-configurational free energy
of the system for each $\gamma$.

It is convenient to relate $F ( N_{\rm A} , N_{\rm X} )$ to
the free energy $F_{\rm A}$ of the pure A crystal having the
same number of lattice sites. The difference
$\Delta F ( N , N_{\rm X} ) \equiv F ( N_{\rm A} , N_{\rm X} ) -
F_{\rm A}$ is the change of free energy when $N_{\rm X}$ atoms of A
in the pure crystal are transmuted into X atoms. We have:
\begin{equation}
\Delta F ( N_{\rm A} , N_{\rm X} ) = - k_{\rm B} T
\ln \left( 
\sum_\gamma e^{- \beta ( \Phi_\gamma - F_{\rm A} )} 
\right) \; .
\end{equation}
Now in the limit $c_{\rm X} \rightarrow 0$, we can neglect the
interactions between X atoms, and we get:
\begin{equation}
\Phi_\gamma - F_{\rm A} \rightarrow N_{\rm X} \mu_{\rm X A}^\dagger \; .
\end{equation}
At higher concentrations, we need to include the free energy
of interaction between pairs of X atoms, and we write:
\begin{equation}
\Phi_\gamma - F_{\rm A} \simeq N_{\rm X} \mu_{\rm X A}^\dagger +
\frac{1}{2} \sum_{m \ne n} \phi_{m n} \; ,
\end{equation}
where $\phi_{m n}$ is the non-configurational free energy change
when a pair of X atoms are brought from widely separated sites 
in the otherwise perfect crystal to
the sites $m$ and $n$. We then have:
\begin{equation}
\Delta F ( N , N_{\rm X} ) = N_{\rm X} \mu_{\rm X A}^\dagger -
k_{\rm B} T \ln \left( \sum_\gamma \exp \left[ - \frac{1}{2} \beta
\sum_{m \ne n} \phi_{m n} \right] \right) \; .
\end{equation}
In the later practical calculations, we approximate by setting
$\phi_{m n}$ equal to zero except when $m$ and $n$ are nearest-neighbor
lattice sites, the interaction free energy being then called
simply $\phi$. 

It is now an exercise in the statistical mechanics of lattice
gases to show that the leading order in $c_{\rm X}$:
\begin{eqnarray}
\Delta F ( N , N_{\rm X} ) = N_{\rm X} \mu_{\rm X A}^\dagger & + &
N k_{\rm B} T \left[ c_{\rm X} \ln c_{\rm X} +
( 1 - c_{\rm X} ) \ln c_{\rm X} \right] + \nonumber \\
& + & \frac{1}{2} N k_{\rm B} T c_{\rm X}^2 z
\left( 1 - e^{- \beta \phi} \right) \; ,
\end{eqnarray}
where $z$ is the coordination number of the lattice. The derivative
$\partial \Delta F ( N , N_{\rm X} ) / \partial N_{\rm X}$
gives us $\mu_{\rm X} - \mu_{\rm A}$, from
which we straightforwardly extract $\lambda_{\rm X}$, which is given by:
\begin{equation}
\lambda_{\rm X} = k_{\rm B} T z \left( 1 - e^{- \beta \phi} \right) \; .
\end{equation}
As in the case of the liquid, this formula should be corrected
from constant volume to constant pressure, so that the correct
formula is:
\begin{equation}
\lambda_{\rm X} = k_{\rm B} T z \left( 1 - e^{- \beta \phi} \right) -
n B_T ( v_{\rm X} - v_{\rm A} )^2 \; ,
\end{equation}
with $B_T$, $v_{\rm X}$ and $v_{\rm A}$ the isothermal bulk
modulus and partial molar volumes in the dilute solid solution.
The calculation of $B_T$ and $v_{\rm A}$ presents no problems;
we return to the {\em ab initio} calculation of $v_{\rm X}$
in Sec.~\ref{sec:partial_molar_volume}.

In addition to using this analytic derivation to obtain
$\lambda_{\rm X}$, we have also performed Monte Carlo
calculations on the lattice gas to obtain numerical
values of $\Delta F ( N , N_{\rm X} )$. These
serve both to confirm the correctness of
the analytic result in the region of low $c_{\rm X}$ and
also to assess deviations from
the linear dependence of ${\tilde{\mu}}_{\rm X}$ on $c_{\rm X}$
as $c_{\rm X}$ increases.

The remaining task is to calculate the nearest-neighbor interaction 
free energy $\phi$. This follows exactly the scheme for
calculating $\mu_{\rm X A}^\dagger$, where now $\phi$ is the
non-configurational free-energy change when two neighboring
atoms in the perfect crystal A are replaced by X atoms, minus twice
$\mu_{\rm X A}^\dagger$. This can be written as:
\begin{equation}
\phi = \phi^{\rm perf} + \phi^{\rm harm} \; ,
\end{equation}
where $\phi^{\rm perf}$ is the (free) energy change for the
perfect non-vibrating crystal, and $\phi^{\rm harm}$ is
the harmonic vibrational part. The static part $\phi^{\rm perf}$
is obtained from the difference of {\em ab initio} free energies
of the relaxed equilibrium system containing neighboring X atoms
and the perfect pure A lattice. We obtain $\phi^{\rm harm}$
from the harmonic vibrational frequencies of the system
containing the neighboring X atoms by a formula exactly
analogous to Eq.~(\ref{eqn:mu_harm}).

We now return very briefly to the question of anharmonicity.
In many cases, very high precision may not be needed for the
chemical potentials, so that anharmonic corrections
to ${\tilde{\mu}}_{\rm X}^s$ can be neglected. But in one
case that will be important later, that of substitutional O
in Fe, we know that anharmonic effects are large. The techniques
we have used to treat them are described in detail 
elsewhere~\cite{alfe00c}.
The strategy is based on thermodynamic integration
between reference models representing both the pure Fe 
and the impure Fe/X systems, followed by further thermodynamic
integrations between the {\em ab initio} and reference systems.

\subsection{Partial molar volumes in the liquid and solid solutions}
\label{sec:partial_molar_volume}

The partial molar volume $v_{\rm X}$ of solute or $v_{\rm A}$ of solvent
is the change of volume of the system when one atom of X or A is added at
constant pressure and temperature. The volumes are related to the
chemical potentials by:
\begin{equation}
v_{\rm X} = ( \partial \mu_{\rm X} / \partial p )_{T , c_{\rm X}} \; ,
\; \; \; \; \; \; \; \; v_{\rm A} = ( \partial \mu_{\rm A} / 
\partial p )_{T , c_{\rm X}} \; .
\label{eqn:partial_molar_volume}
\end{equation}
We note that the total volume of the system is given by $V = N_{\rm X}
v_{\rm X} + N_{\rm A} v_{\rm A}$. As for the chemical potentials, we
find it easier to consider the interconversion of solvent and
solute, and to work with the difference $v_{\rm X A} \equiv v_{\rm
X} - v_{\rm A}$. The liquid is treated by {\em ab initio} m.d., in
which the pressure for a given volume is calculated during the
simulation. (In our practical calculations, we work at constant
$V$.) The straightforward way of obtaining the dilute limit of
$v_{\rm X}$ is therefore to calculate the change of pressure $\Delta
p$ resulting from the replacement of a chosen number $N_{\rm X}$
of atoms in the pure solvent by X. The pressure change per atom
$\delta p = \Delta p / N_{\rm X}$ then gives us $v_{\rm X A}$ by the
relation $v_{\rm X A} = V \delta p / B_T$. 

It is clearly possible
to follow the same route for the solid. However, if the solid is
treated by harmonic frequency calculations with or without
thermodynamic integration for the anharmonic contribution, then the
partial molar volumes must be obtained from the chemical potentials
{\em via} Eq.~(\ref{eqn:partial_molar_volume}). This requires
calculation of $\mu_{\rm X}$ at different volumes followed by
numerical differentiation.

\section{Illustration: chemical equilibrium in the Earth's core}
\label{sec:earths_core}

In applying the techniques to study chemical phase equilibrium
between the Earth's inner and outer core, our aim is to
show how they can yield important new information about the chemical
composition and temperature of the core, both of which are controversial.
Our strategy exploits the fact that that the density as a function
of depth in the core is accurately known from seismic
measurements~\cite{prem}; in particular, it is quite well established that
there is a density discontinuity of $4.5 \pm 0.5$~\% across the 
inner-core/outer-core boundary (ICB)~\cite{shearer90}. Recent {\em ab initio}
studies of the melting properties of pure Fe concur in giving
a volume of fusion of $\sim 1.8$~\%~\cite{alfe01c,laio00}, 
which is clearly much smaller.
This means that
there must be a substantial partitioning of light solute elements
from solid to liquid to account for the large observed discontinuity.

This discontinuity can be studied with our methods. If we suppose
initially that the core is a binary solution of Fe with one of the
leading impurity condidates S, Si or O~\cite{poirier94}, 
then the solute mole fraction
in the liquid core can be fixed by requiring that the density
reproduce the seismically observed density.  Calculation of the
chemical potentials $\mu_{\rm X}$ in the liquid and solid then gives
us the mole fraction in the solid, from which we can deduce the solid
density, and hence the density discontinuity. Agreement or
disagreement with the known discontinuity puts a constraint on the
composition.  At the same time, the shift of melting temperature given
by Eq.~(\ref{eqn:Tm_shift}) gives us information about the temperature
at the ICB.

In the following, we first summarize the general techniques used
in all the calculations (Sec.~\ref{sec:general_techniques}). 
We then describe separately the calculations
on the liquid and solid alloys (Sec.~\ref{sec:results_liquid}
and \ref{sec:results_solid} respectively),
presenting results for the chemical potentials and partial molar
volumes. In Sec.~\ref{sec:results_core}, we then 
combine the results with seismic data
to obtain constraints on the chemical composition and temperature
of the Earth's core.

\subsection{General techniques}
\label{sec:general_techniques}

Our {\em ab initio} calculations are based on the well established DFT
methods used in virtually all {\em ab initio} investigations of solid
and liquid
Fe~\cite{alfe01a,alfe99a,alfe01c,stixrude94,soderlind96,dewijs98a,alfe00a}, 
including our own previous work on pure Fe and its solid and liquid
alloys with S and O~\cite{alfe00b,alfe00c,alfe98,alfe99c}. We
employ the generalized gradient approximation for exchange-correlation
energy, as formulated by Perdew {\em et al.}~\cite{perdew92}, which is
known to give very accurate results for the low-pressure elastic,
vibrational and magnetic properties of body-centred cubic (b.c.c.) Fe,
the b.c.c.~$\rightarrow$ h.c.p. transition pressure, and the
pressure-volume relation for h.c.p. Fe up to over
300~GPa~\cite{stixrude94,alfe00a}. There is also very recent
evidence for their accuracy in predicting the high-pressure
phonon spectrum of h.c.p. Fe~\cite{mao01}.
We use the ultra-soft
pseudopotential implementation~\cite{vanderbilt90} of DFT with
plane-wave basis sets, an approach which has been demonstrated to give
results for solid Fe that are virtually identical to those of
all-electron DFT methods~\cite{alfe00a}.  Our calculations are
performed using the VASP code~\cite{kresse96}, which is exceptionally
stable and efficient for metals. We implemented a scheme for the
extrapolation of the charge density which increases the efficiency of
the molecular dynamics simulations by nearly a factor of
two~\cite{alfe99b}.  The technical details of pseudopotentials,
plane-wave cut-offs, etc.  are the same as in our previous
work~\cite{alfe99c}.

\subsection{The liquid}
\label{sec:results_liquid}
Our {\em ab initio} m.d. simulations on the liquid, which we used to
calculate $W ( N , N_{\rm X} )$ and hence the chemical potentials,
were all performed on systems of 64 atoms, with a time-step of 1~fs
and with $\Gamma$-point sampling of the electronic Brillouin zone. In
our previous calculations on pure liquid Fe~\cite{alfe01c}, we showed
that $\Gamma$-point sampling on a 67-atom cell underestimates the free
energy by {\em ca.}~10~meV/atom; this is a completely negligible error
for present purposes. The calculations were done at $T = 7000$~K and
at the volume/atom $V / N = 6.97$~\AA/atom, which for pure Fe gives a
pressure of 370~GPa. This pressure is somewhat higher
than the ICB pressure of 330~GPa~\cite{prem}. The temperature is also
higher than that at the ICB: our {\em ab initio} melting curve gives a
melting temperature of $\sim 6350$~K (or $\sim 6200$~K after the
correction due to our estimate of likely DFT errors)~\cite{alfe01c} at the ICB
pressure of 330~GPa, which is already higher than some other
estimates~\cite{laio00,anderson97}. But we shall see below that
depression of freezing point due to impurity partitioning lowers this
by a further $\sim 700$~K. We have made rough estimates which show
that the difference between 7000~K and our estimated ICB temperature
is unlikely to change the chemical potentials of S and Si by more than
0.1~eV and that of O by more than 0.3~eV, which will have no
significant effect on our conclusions. The difference of pressures
should also make little difference.

\begin{figure}
\psfig{figure=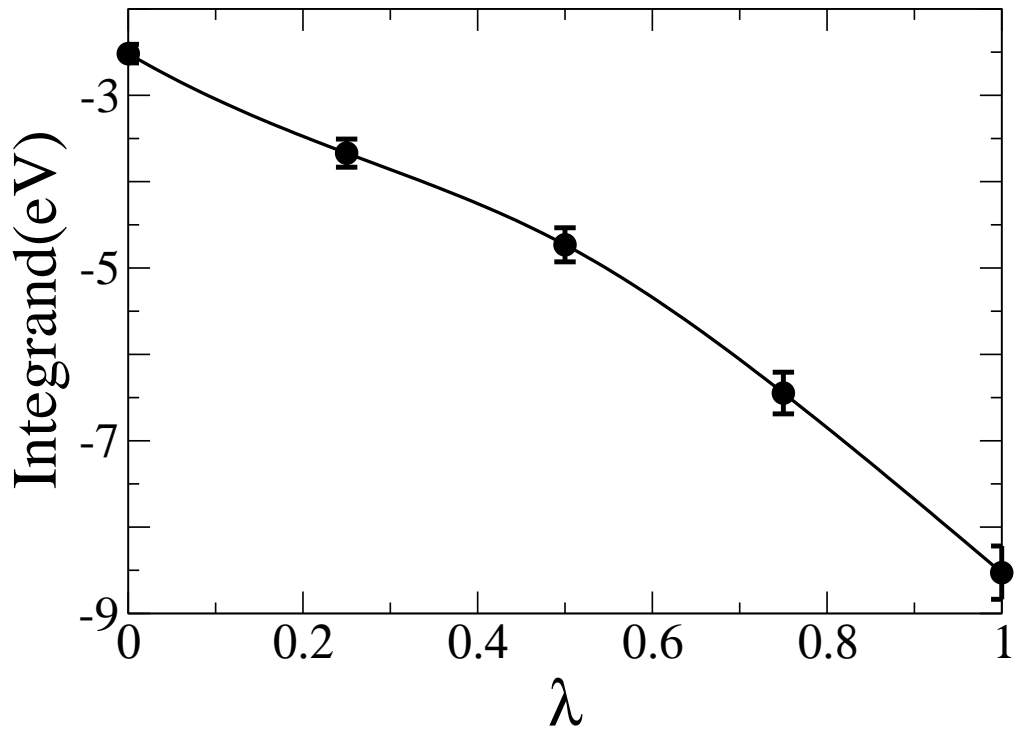,height=4.5in}
\caption{The integrand $\langle U_1 - U_0 \rangle_\lambda$ (eV units)
appearing in the thermodynamic integration used to calculate
the free energy change $W ( N , N_{\rm X} )$ when $N_{\rm X}$
atoms of pure solvent are converted into solute
atoms, with total number of atoms in the system $= N$ (see
Eq.~(\ref{eqn:thermint_W})). Results shown refer to oxygen solute
for $N_{\rm X} = 12$ and $N = 64$. Filled circles show values computed from
{\em ab initio} m.d. simulations, with bars indicating statistical
errors. Curve is a polynomial fit to the computed values.}
\label{fig:lambda}
\end{figure}

\begin{figure}
\psfig{figure=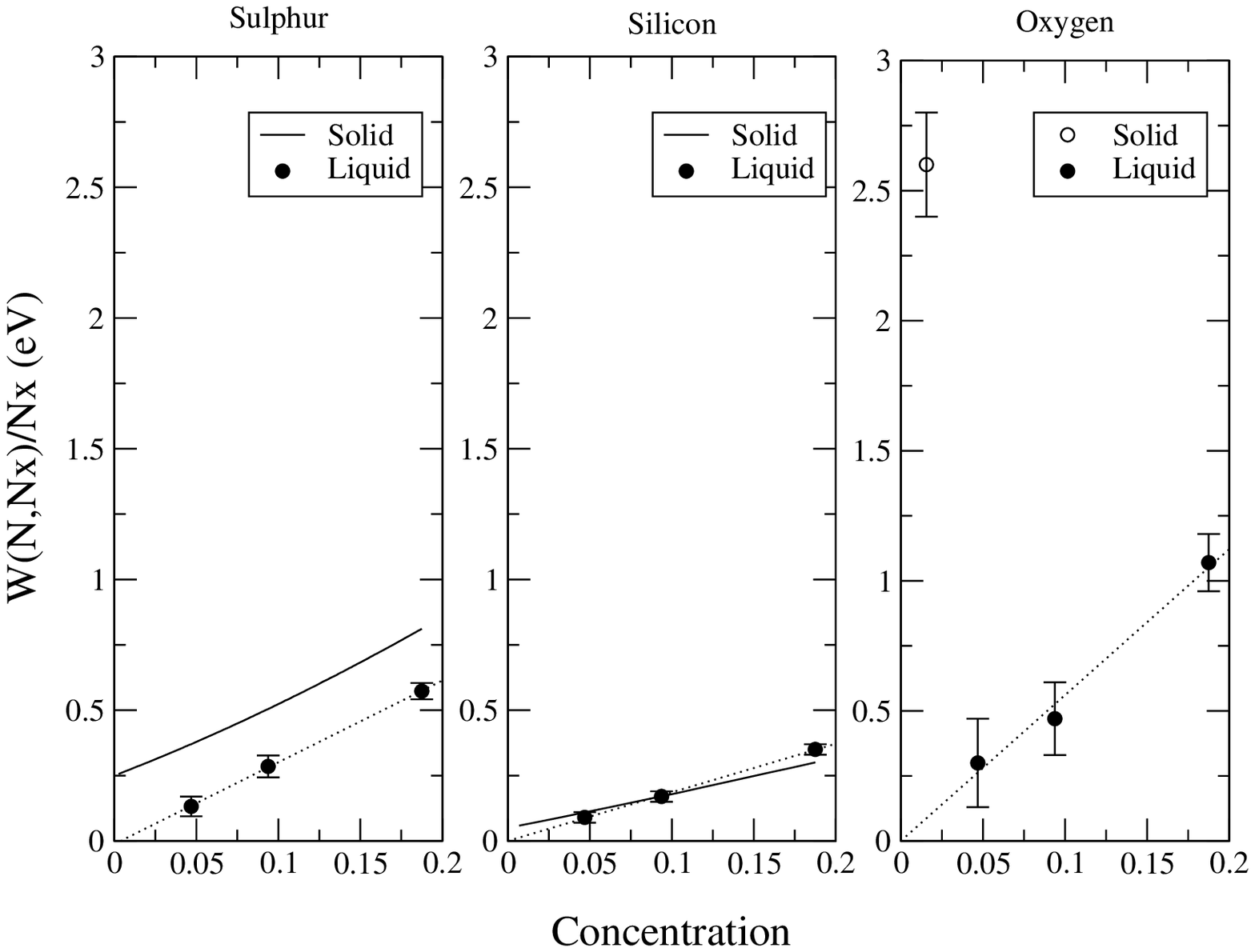,height=4.5in}
\caption{The calculated free energy change $W ( N , N_{\rm X} )$ when
$N_{\rm X}$ atoms of pure solvent are converted into solute
atoms, with total number of atoms in the system $= N$. Quantity
plotted is $W ( N , N_{\rm X} ) / N_{\rm X}$ (eV units) as
a function of concentration $c_{\rm X} = N_{\rm X} / N$ for
liquid and solid solutions of S, Si and O in Fe. Filled circles
are results for liquid, with bars indicating statistical
errors, and straight dotted line being a least-squares fit to
these data. Continuous curves for S and Si show results for solid solution
obtained from Monte Carlo calculations based on {\em ab initio}
free energy of nearest-neighbor interaction. Open circle with error bar
is result for O in solid from thermodynamic integration.}
\label{fig:W}
\end{figure}

We have used thermodynamic integration to 
calculate $W ( N , N_{\rm X} )$ for the three solute 
elements S, Si and O for $N_{\rm X} = 3$, 6 and 12, corresponding 
to mole fractions of 4.7, 9.4 and 18.8~\%. In doing this,
we have aimed to choose the number of $\lambda$ values large enough
and the duration of the simulation at each $\lambda$ value long
enough to give a precision on $W ( N , N_{\rm X} ) / N_{\rm X}$
of {\em ca.}~0.05~eV for S and Si and {\em ca.}~0.1~eV for O.
To illustrate how the thermal average $\langle U_1 - U_0 \rangle_\lambda$
in Eq. (\ref{eqn:thermint_W}) depends on $\lambda$,
we display this quantity in Fig.~\ref{fig:lambda} at five
equally spaced $\lambda$ values for the oxygen system
with $N_{\rm X} = 12$. We see that the dependence on $\lambda$
is not far from linear. Using Simpson's rule to perform the
integral, we compared results for $W ( N , N_{\rm X} ) / N_{\rm X}$
using the five $\lambda$ values 0.0, 0.25, 0.50, 0.75 and 1.00 with those
obtained using only the three values 0.0, 0.5 and 1.0, and found
that they differ by less than the statistical error.
Since the replacement of Fe by O is a greater
perturbation than that of Fe by S or Si (see below), we have taken this
as justification for using only three $\lambda$ values in all
the thermodynamic integrations.
Our numerical results for $W ( N , N_{\rm X} ) /
N_{\rm X}$ for X = S, Si and O, together with the linear
least-square fit of Eq.~(\ref{eqn:w}), are reported in Fig.~\ref{fig:W},
and the resulting values of $a \equiv \mu_{\rm XA}^\dagger$ and $b$ are
given in Table~\ref{tab:chem_pot}. 

As explained in
Sec.~\ref{sec:liqsol}, the $b$ values have to be corrected as in
Eq.~(\ref{eqn:lambda_correction}) in order to obtain 
$\lambda_{\rm X}$, and this requires the partial molar volumes 
$v_{\rm X}$. We
obtain the partial molar volumes from the simulations just described
by studying the pressure change resulting from the replacement of
$N_{\rm X}$ atoms in the pure liquid by atoms of X at constant volume
-- this is straightforward, since the pressure is automatically
calculated during the constant-volume simulations. We find that
within the statistical errors the change of pressure is linear in
$c_{\rm X}$ for all three impurity species. We then use the fact that
$v_{\rm X} - v_{\rm A} = ( \bar{v} / B_T ) ( \partial p / \partial
c_{\rm X} )_T$; for $B_T$ we use the bulk modulus of the pure
liquid, which we know from our previous 
work~\cite{alfe01c}. The calculated $v_{\rm
X}$ values are 6.65, 6.65 and 4.25~\AA$^3$ for S, Si and O
respectively, compared with the volume per atom in the pure liquid
of 6.97~\AA$^3$. We note that S and Si have almost exactly the same
volume as Fe, but that the volume of O is considerably smaller.
Using these $v_{\rm X}$ values in Eq.~(\ref{eqn:lambda_correction}),
we now obtain the results for $\lambda_{\rm X}$ given in
Table~\ref{tab:chem_pot}. We see that the difference between $2 b$ and
$\lambda_{\rm X}$ is very small for sulfur and silicon, as expected,
but is substantial for oxygen.

\subsection{The solid}
\label{sec:results_solid}

The available evidence strongly indicates that the stable
crystal structure of Fe at the pressures and temperatures of
the Earth's core is hexagonal close packed (h.c.p.)~\cite{vocadlo00},
and this is the structure adopted in our calculations. We first
present our calculations for S and Si, and then summarize briefly our
results for the more complex case of O, which have already been
reported elsewhere.


\subsubsection{Sulfur and silicon}

The calculations are performed on a $4\times 4\times 2$
h.c.p. supercell containing 64 atoms, with a $3\times 3\times 2$
Monkhorst-Pack~\cite{monkhorst76} grid of 
electronic $k$-points which give
free energies converged within a few meV/atom. In our calculations
on the static zero-temperature lattice, we find that when a single
Fe is replaced by S or Si at constant volume, the relaxation of
neighboring atoms is very small, and the pressure change
is also small. For the atomic volume $v = 6.97$~\AA$^3$/atom,
the partial molar volumes calculated without lattice vibrations
give the differences
$v_{\rm S} - v_{\rm Fe} = -0.32$ and $v_{\rm Si} - v_{\rm Fe} = -0.32$~\AA$^3$,
which are extremely small compared with $v_{\rm Fe}$. We assume that
that these differences will not be significantly affected by
thermal effects.

We now turn to the harmonic frequencies $\omega_n$ and
$\omega_n^\prime$ needed for the harmonic difference of chemical
potentials $\mu_{\rm XA}^{\dagger {\rm harm}}$ (see
Eq.~(\ref{eqn:mu_harm})).  We calculate these using a supercell of 64
atoms in all calculations. For the pure Fe system only two independent
displacements of a single atom are needed to obtain the full force
constant matrix~\cite{alfe01a}. We displace the atom by {\em
ca.}~0.015~\AA\ in each direction, which is known to be small enough
to ensure accurate linearity between forces and displacements.  For
the calculations where one Fe is substituted with S or Si, the
symmetry is much reduced, and the number of atoms to be displaced is
15, with the total number of independent displacements being 33. For
the systems with two solute atoms, the symmetry of the system is
reduced even further, and we need to displace 20 atoms in all possible
directions, for a total of 60 displacements. There are two distinct
ways of putting two S or Si atoms on nearest-neighbor sites: the
first has both sites in the same basal plane, and the second has them
in adjacent basal places. Within our errors, we cannot detect the free
energy difference between the two arrangements. Since the zero
temperature value of the difference $v_{\rm X} - v_{\rm Fe}$ is very
small, we have not attempted to calculate its high temperature value in
the harmonic approximation, and we report in Table~\ref{tab:chem_pot}
the zero temperature value. The correction to $\lambda_{\rm X}$ is
negligible anyway and can be ignored.

For sulfur and silicon we neglect anharmonic corrections. In our
previous work on pure Fe~\cite{alfe01a} we showed that at ICB
conditions the anharmonic contribution to the free energy is roughly
60 meV/atom. In this case we are concerned with free energy
differences between the pure Fe system and a system where one of the
Fe atoms has been substituted with X, so the difference of the
relative anharmonic contributions to the free energies is presumably
smaller than that. 

Our calculated values of $\mu_{\rm XA}^{\dagger s}$ and 
$\lambda_{\rm X}^s$ at $v = 6.97$~\AA$^3$/atom and $T = 7000$~K
are reported in Table~\ref{tab:chem_pot}.

\subsubsection{Oxygen}

As emphasized above and in previous work~\cite{alfe00c}, 
substitutional O in h.c.p. Fe is
highly anharmonic, because O is considerably smaller than Fe and
has great freedom of movement, so that the harmonic approximation
is completely inadequate for calculating 
$\mu_{{\rm O} \, {\rm Fe}}^{\dagger s}$.
We gave a brief summary in Sec.~\ref{sec:solsol} of the thermodynamic
integration techniques used to do the calculations. The numerical
result for $\mu_{{\rm O} \, {\rm Fe}}^{\dagger s}$ at 
$v = 6.97$~\AA$^3$/atom and $T = 7000$~K is reported in 
Table~\ref{tab:chem_pot}.
We have not attempted to calculate $\lambda_{\rm X}$ for X~=~O,
since this would be extremely demanding, and turns out to be
unnecessary in our analysis of core composition.

To calculate $v_{\rm O} - v_{\rm Fe}$ in the solid we have repeated
the calculations at different volumes and numerically differentiated
the results, as described in
section~\ref{sec:partial_molar_volume}. The value of $v_{\rm O} -
v_{\rm Fe}$ is reported in Table~\ref{tab:chem_pot}.

\subsection{Core composition and temperature}
\label{sec:results_core}

Some crucial features of our results are immediately clear from
Table~\ref{tab:chem_pot}: the liquid-solid difference $\mu_{\rm
XA}^{\dagger l} - \mu_{\rm XA}^{\dagger s}$ is negative in all cases;
its magnitude is somewhat smaller than $k_{\rm B} T$ for S and Si, but
is much bigger than $k_{\rm B} T$ for O. This implies that the solutes
will all partition from solid into liquid, as expected; but the
partitioning will be weak for S and Si and very strong for O.

To see the implications in more detail, consider the case of Fe/S.  If
we postulate that the core is an Fe/S binary alloy, then we can
estimate the mole fraction $c_{\rm S}^l$ in the outer core by noting
that the density of pure liquid iron at the ICB pressure
is {\em ca.}~6~\% higher than the values obtained from seismic
data~\cite{prem}. We therefore add sulfur to the liquid until the density is 
reduced to the required value, which gives $c_{\rm S}^l = 0.16$.
Now if we ignored the dependence of
${\tilde{\mu}}_{\rm S}$ on concentration, then the value $\mu_{\rm
S}^{\dagger l} - \mu_{\rm S}^{\dagger s} = -0.25$~eV would give
$c_{\rm S}^s / c_{\rm S}^l = 0.66$, so that $c_{\rm S}^s = 0.11$.
However, the positive $\lambda_{\rm S}$ values mean that both
${\tilde{\mu}}_{\rm S}^l$ and ${\tilde{\mu}}_{\rm S}^s$ increase
strongly with increasing mole fraction of S, and this will tend to
equalize the mole fractions in solid and liquid. If we solve
Eq.~(\ref{eqn:partition}) self-consistently for $c_{\rm S}^s$ with the
given $c_{\rm S}^l$, we find $c_{\rm S}^s = 0.14$. But a 14~\% mole
fraction of S in the inner core is completely incompatible with the
seismic measurements. We can use the $c_{\rm S}^s$ volume together
with the partial molar volume $v_{\rm S}^s$ to calculate the change of
density of the solid due to dissolved S, and hence the ICB density
discontinuity.  We find the discontinuity is increased from the
pure-Fe value of 1.8~\% up to $2.7 \pm 0.5$~\%, which is still much
less than the seismic value of $4.5 \pm 0.5$~\%. This means that the
binary Fe/S alloy can be ruled out as a model for core composition.
The argument is still stronger for Si, since the chemical 
potentials in solid and liquid are even more similar than for S.
We conclude that the binary Fe/Si model must also be ruled out.

For O, the situation is the opposite. The difference of chemical
potentials in liquid and solid has the very large value $\mu_{{\rm O}
\, {\rm Fe}}^{\dagger l} - \mu_{{\rm O} \, {\rm Fe}}^{\dagger s} = -
2.6$~eV, which imples a strong partitioning from solid to liquid. If
we repeat our analysis of the outer-core density with the partial
molar volume $v_{\rm O}^l$, we find that an oxygen mole fraction
$c_{\rm O}^l = 0.18$ is needed to match the density of the outer
core. Eq.~(\ref{eqn:partition}) then gives $c_{\rm O}^s \simeq 0.003$,
so that the O concentration in the inner core is very small.  With our
calculated $v_{\rm O}^s$ value, we then find an ICB density
discontinuity of $7.8 \pm 0.2$~\%, which is markedly larger than the
seismic value. A binary Fe/O model can thus also be ruled out.

Although all the binary models fail, the seismic data can clearly be
accounted for by ternary or quaternary alloys of the three
impurities. {\em Ab initio} calculations on such liquid and solid
alloys would certainly be feasible with the methods we have developed,
but would need a considerably greater effort. If we assume for the
moment that the different impurities do not affect each other's
chemical potentials, we can use our present results to construct a
model for the core composition.  We have seen that S and Si alone
cannot explain the density jump at ICB, so there must be some O in
the outer core. If we dissolve some O in liquid iron, together with
S/Si, maintaining the density of the alloy equal to the density of the
core, we increase the density jump at the ICB. This is because hardly
any O goes into the solid. We therefore continue to add O till we
match the density jump at ICB. 
The resulting chemical compositions of the inner and
outer core are summarized in Table~\ref{tab:mol_frac}.

With these compositions, Eq.~(\ref{eqn:partition}) now allows us to
determine the shift of melting temperature from that of pure Fe at the
ICB pressure; we find $\Delta T_m = - 700 \pm 100$~K.  Comparing this
with our earlier {\em ab initio} melting temperature $T_m = 6200-6350$~K
for pure Fe at $p = 330$~GPa~\cite{alfe01c}, we 
obtain the estimate $T_{\rm ICB} \sim
5600$~K for the temperature at the boundary between inner and outer
core. This is quite close to estimates that have been made in other 
ways~\cite{anderson97}. The implications of our temperature and
chemical composition results for the understanding of the Earth's
dynamics and past history will be explored elsewhere.

\section{Discussion and conclusions}
\label{sec:discuss}

We have shown the practical feasibility of calculating
completely {\em ab initio} chemical potentials
in liquids and solids, and hence the {\em ab initio}
treatment of chemical equilibrium between coexisting phases.
The practical benefits of being able
to do such calculations have also been illustrated by
showing how they can help to improve our understanding
of a controversial and important chemical-equilibrium
problem in the earth sciences. We note that, although the calculations
are demanding at present because of the need to perform
substantial {\em ab initio} molecular dynamics simulations,
the underlying concepts are rather straightforward,
and represent a simple extension of well-known classical techniques.

In conclusion, we want to stress that the techniques should have
rather wide applications. Although we have chosen to focus on
the partitioning of impurities between coexisting solid and liquid phases,
the methods could equally well be used to study partitioning between
liquid phases, or between solid phases. The ability to calculate
{\em ab initio} chemical potentials in liquids also makes
it possible to contemplate the {\em ab initio}
calculation of the solubility of solids, liquids or gases in liquids.
The practical application of these ideas is likely to
be limited only by the need to find economical thermodynamic
integration paths for transforming chemical species into each other.

\section*{Acknowledgments}

The work of DA is supported by NERC grant GST/02/1454 to G. D. Price
and M. J. Gillan and by a Royal Society University Research
Fellowship. We thank NERC and EPSRC for allocations of time on the
Cray T3E machines at Edinburgh Parallel Computer Centre and Manchester
CSAR service, these allocations being provided through the Minerals
Physics Consortium (GST/02/1002) and the UK Car-Parrinello Consortium
(GR/M01753). We also acknowledge use of facilities at the UCL
HiPerSpace Centre, funded by the Joint Research Equipment Initiative.
We gratefully acknowledge discussions with Dr. L. Vo\v{c}adlo.

\setcounter{equation}{0}
\def\theequation{A\arabic{equation}}

\section*{Appendix: from constant volume to constant pressure}
\label{sec:appendix}

We explained in the text that the isobaric dependence of solute
chemical potential on concentration is obtained from
$( \partial m / \partial c_{\rm X} )_p$, where $m$ is the
non-trivial part of the solute chemical potential, defined in
Eq.~(\ref{eqn:m}). However, the quantity given by our {\em ab initio}
calculations is $( \partial m / \partial c_{\rm X} )_V$. We derive
here the relation between the isobaric and isochoric derivatives of $m$.

We start by noting that:
\begin{eqnarray}
( \partial m / \partial c_{\rm X} )_p & = &
( \partial m / \partial c_{\rm X} )_V +
( \partial m / \partial V )_{c_{\rm X}} 
( \partial V / \partial c_{\rm X} )_p \nonumber \\
& = & ( \partial m / \partial c_{\rm X} )_V +
( \partial m / \partial V)_{c_{\rm X}}
N ( v_{\rm X} - v_{\rm A} ) \; ,
\label{eqn:a1}
\end{eqnarray}
with $T$ held constant throughout, where $N$ is the total number
of atoms, and we have used the basic definition of the partial
molar volumes $v_{\rm X}$ and $v_{\rm A}$ of solute and solvent.
Next, we refer to Eq.~(\ref{eqn:mu_xa2}) to see that:
\begin{equation}
( \partial m / \partial V )_{c_{\rm X}} =
( \partial ( \mu_{\rm X} - \mu_{\rm A} ) /
\partial V )_{c_{\rm X}} \; ,
\label{eqn:a2}
\end{equation}
which can be reexpressed as:
\begin{equation}
( \partial m / \partial V )_{c_{\rm X}} =
( \partial ( \mu_{\rm X} - \mu_{\rm A} ) /
\partial p )_{c_{\rm X}} ( \partial p / \partial V )_{c_{\rm X}} 
= - ( v_{\rm X} - v_{\rm A} ) B_T / V \; ,
\label{eqn:a3}
\end{equation}
with $B_T$ the isothermal bulk modulus, and we have used
the relations $( \partial \mu_{\rm X} / \partial p )_{c_{\rm X}} =
v_{\rm X}$ and $( \partial \mu_{\rm A} / \partial p )_{c_{\rm A}} =
v_{\rm A}$. Combining Eqs.~(\ref{eqn:a1}) and (\ref{eqn:a3}), we have:
\begin{equation}
( \partial m / \partial c_{\rm X} )_p =
( \partial m / \partial c_{\rm X} )_V -
n B_T ( v_{\rm X} - v_{\rm A} )^2 \; ,
\label{eqn:a4}
\end{equation}
where $n = N / V$ is the overall atomic number density.

\newpage
\pagebreak

\pagebreak

\begin{table}
\begin{tabular}{l|ccc}
Solute & S & Si & O \\
\hline
$\mu_{\rm XA}^{\dagger l}$ & $3.5 \pm 0.05$ &  $2.35 \pm 0.02$ & $-6.25 \pm 0.2$ \\
$b^l_{\rm X}$              & 3.13 & 1.86 & 5.6 \\
$v^l_{\rm XA}$             & -0.32 & -0.32 & -2.72 \\
$\lambda^l_{\rm X}$          & 6.15 & 3.6 & 3.25 \\
$\mu_{\rm XA}^{\dagger s}$ & $3.75 \pm 0.05$ & $2.40 \pm 0.02$ & $-3.65 \pm 0.2$\\
$b^s_{\rm X}$              & 3.0  & 1.4 & \\
$v^s_{\rm XA}$             & -0.32 & -0.32 & -2.35\\
$\lambda^s_{\rm X}$        & 5.9 & 2.7 & \\
$\mu_{\rm XA}^{\dagger l} - \mu_{\rm XA}^{\dagger s}$ & $-0.25 \pm 0.04$ & $-0.05 \pm 0.02$ & $-2.6 \pm 0.2$ \\
\hline
\end{tabular}
\caption{Calculated chemical potentials (eV units) and partial
atomic volumes $v_{\rm X}$ (\AA$^3$ units) of solutes 
X~= S, Si and O in liquid and
h.c.p. solid Fe at conditions close to those of the Earth's core
(see text). Chemical potential of X is represented at low mole
fraction $c_{\rm X}$ by $\mu_{\rm X} = k_{\rm B} T \ln c_{\rm X} +
{\tilde{\mu}}_{\rm X}$, with ${\tilde{\mu}}_{\rm X}$ linearized
as ${\tilde{\mu}}_{\rm X} \simeq \mu_{\rm X}^\dagger + 
\lambda_{\rm X} c_{\rm X}$. The quantity $\mu_{X A}^\dagger$ is
$\mu_{\rm X}^\dagger - \mu_{\rm Fe}^0$, with $\mu_{\rm Fe}^0$
the chemical potential of pure solvent Fe; $v_{\rm X A}$ is
$v_{\rm X} - v_{\rm Fe}$, with $v_{\rm Fe}$ the volume per atom
in pure Fe. The meaning of the calculated quantity $b_{\rm X}$
used to obtain $\lambda_{\rm X}$ is explained in Sec.~\ref{sec:liqsol}.
Superscripts $l$ and $s$ indicate liquid and solid.}
\label{tab:chem_pot}
\end{table}

\begin{table}
\begin{tabular}{lcc}
\hline
& Solid & Liquid \\
Sulfur/Silicon & $8.5 \pm 2.5 $ & $10 \pm 2.5 $  \\
Oxygen & $0.2 \pm 0.1 $ & $8.0 \pm 2.5 $  \\
\hline
\end{tabular}
\caption{Estimated molar percentages of sulfur, silicon and oxygen
in the Earth's solid inner core and liquid outer core obtained by combining
{\em ab initio} calculations and seismic data. Sulfur/silicon entries
refer to total percentages of sulfur and/or silicon.}
\label{tab:mol_frac}
\end{table}

\pagebreak

\begin{thebibliography}{99}

\bibitem{generaldft} P. Hohenberg and W. Kohn, Phys. Rev. {\bf 136},
B864 (1964); W. Kohn and L. Sham, Phys. Rev. {\bf 140}, A1133 (1965);
R. O. Jones and O. Gunnarsson, Rev. Mod. Phys. {\bf 61}, 689 (1989);
M. C. Payne, M. P. Teter, D. C. Allan, T. A. Arias, and
J. D. Joannopoulos, Rev. Mod. Phys. {\bf 64}, 1045 (1992);
M. J. Gillan, Contemp. Phys. {\bf 38}, 115 (1997); R. G. Parr and
W. Yang, {\em Density-Functional Theory of Atoms and Molecules},
(Oxford University Press, Oxford, 1989).

\bibitem{baroni87}
S. Baroni, P. Giannozzi, and A. Testa, Phys. Rev. Lett. {\bf 58}, 1861 (1987).

\bibitem{karki00} B. B. Karki, R. M. Wentzcovitch, S. de Gironcoli, and
S. Baroni, Phys. Rev. B {\bf 62}, 14750 (2000).

\bibitem{Lichtenstein00} A. I. Lichtenstein, R. O. Jones, S. de
Gironcoli, and S. Baroni, Phys. Rev. B {\bf 62},  11487 (2000).

\bibitem{xie99a}
J. J. Xie, S. P. Chen, J. S. Tse, S. de Gironcoli, and S.  Baroni,
Phys. Rev. B, {\bf 60}, 9444 (1999).

\bibitem{xie99b}
J. J. Xie, S. de Gironcoli, S.  Baroni, and M. Scheffler,
Phys. Rev. B, {\bf 59}, 965 (1999).

\bibitem{lazzeri98}
M. Lazzeri and S. de Gironcoli, Phys. Rev. Lett. {\bf 81}, 2096 (1998).

\bibitem{christensen00a}  
N. E. Christensen, Phys. Stat. Sol. B {\bf 220}, 325 (2000).

\bibitem{alfe01a}
D. Alf\`e, G. D. Price, and M. J. Gillan, Phys. Rev. B {\bf 64}, 045123 (2001).

\bibitem{pavone98}       
P. Pavone, S. de Gironcoli, and S. Baroni,  Phys. Rev. B 
{\bf 57}, 10421 (1998).

\bibitem{kern99}	
G. Kern, G. Kresse, and J. Hafner, Phys. Rev. B {\bf 59},
8551 (1999).

\bibitem{gaal-nagy99}    
K. Gaal-Nagy, A. Bauer, M. Schmitt, K. Karch,
P. Pavone, and D. Strauch, Phys. Stat. Sol. B {\bf 221}, 275 (1999).

\bibitem{christensen00b}  
N. E. Christensen, D. J. Boers, J. L. van Velsen, and D. L. Novikov,
J. Phys. Cond. Matt. {\bf 12}, 3293 (2000).

\bibitem{car85} R. Car and M. Parrinello, Phys. Rev. Lett. {\bf 55},
2471 (1985).

\bibitem{sugino95} O. Sugino and R. Car, Phys. Rev. Lett. {\bf 74},
1823 (1995).

\bibitem{dewijs98} G. A. de Wijs, G. Kresse, and M. J. Gillan, 
Phys. Rev. B {\bf 57}, 8223 (1998).

\bibitem{alfe99a} D. Alf\`{e}, M. J. Gillan, and G. D. Price,
Nature {\bf 401}, 462 (1999).


\bibitem{alfe01c} D. Alf\`{e}, M. J. Gillan, and G. D. Price,
Phys. Rev. B, submitted. Preprint available in electronic form
at http://arXiv.org/ps/cond-mat/0107307.

\bibitem{vocadlo01} L. Vo\v{c}adlo and D. Alf\`e, Phys. Rev. B,
submitted. Preprint available in electronic form at
http://arXiv.org/ps/cond-mat/0108460.

\bibitem{jesson2000} B. J. Jesson and P. A. Madden, 
J. Chem. Phys. {\bf 113} 5924 (2000).

\bibitem{smargiassi96}
E. Smargiassi and R. Car, Phys. Rev. B {\bf 53}, 9760 (1996).

\bibitem{smargiassi95} E. Smargiassi and P. A. Madden, Phys. Rev. B
{\bf 51}, 117 (1995).

\bibitem{arthur98}	
J. W. Arthur and A. D. J. Haymet, J. Chem. Phys. {\bf 109},
7991 (1998).

\bibitem{birch64} F. Birch, J. Geophys. Res. {\bf 69}, 4377 (1964).

\bibitem{ringwood77} A. E. Ringwood, Geochem. J. {\bf 11}, 111
(1977).

\bibitem{poirier94} J.-P. Poirier, Phys. Earth Planet. Inter.
{\bf 85}, 319 (1994).

\bibitem{alfe00b} D. Alf\`{e}, M. J. Gillan, and G. D. Price, Nature
{\bf 405} 172 (2000).
 
\bibitem{laio00}
A. Laio, S. Bernard, G. L. Chiarotti, S. Scandolo, and E. Tosatti,
Science {\bf 287}, 1027 (2000).

\bibitem{shearer90} P. Shearer and G. Masters, Geophys. J. Int.
{\bf 102}, 491 (1990); T. G. Masters and P. M. Shearer,
J. Geophys. Res. {\bf 95}, 21691 (1990).

\bibitem{prem} A. M. Dziewonski and D. L. Anderson, Phys. Earth
Planet. Inter. {\bf 25}, 297 (1981).

\bibitem{alfe00c} D. Alf\`{e}, M. J. Gillan, and G. D. Price,
Geophys. Res. Lett. {\bf 27}, 2417 (2000).

\bibitem{alfe01b} D. Alf\`{e}, M. J. Gillan, and G. D. Price, 
Earth Planet. Sci. Lett., submitted.

\bibitem{frenkel96} D. Frenkel and B. Smit, {\em Understanding
Molecular Simulation}, Ch. 4 (Academic, New York, 1996).

\bibitem{kresse95} G. Kresse, J.  Furthm\"{u}ller and J. Hafner, 
Europhys. Lett. {\bf 32}, 729 (1995).

\bibitem{darioweb} Program available at {\tt
http://chianti.geol.ucl.ac.uk/$\sim$dario}

\bibitem{stixrude94} L. Stixrude, R. E. Cohen, and D. J. Singh, 
Phys. Rev. B, {\bf 50}, 6442 (1994).

\bibitem{soderlind96} P. S\"{o}derlind, J. A. Moriarty, and
J. M. Wills, Phys. Rev. B {\bf 53}, 14063 (1996).

\bibitem{dewijs98a} G. A. de Wijs, G. Kresse, L. Vo\v{c}adlo,
D. Dobson, D. Alf\`{e}, M. J. Gillan, and G. D. Price, Nature
{\bf 392}, 805--807 (1998).

\bibitem{alfe00a} D. Alf\`{e}, G. Kresse, and M. J. Gillan, 
Phys. Rev. B {\bf 61}, 132 (2000).

\bibitem{alfe98} D. Alf\`{e} and M. J. Gillan,  Phys. Rev. B 
{\bf 58}, 8248 (1998).

\bibitem{alfe99c} D. Alf\`{e}, G. D. Price, and M. J. Gillan, 
Phys. Earth Planet. Inter. {\bf 110}, 191 (1999).

\bibitem{perdew92}
J. P. Perdew, J. A. Chevary, S. H. Vosko, K. A. Jackson,
M. R. Pederson, D. J. Singh, and C. Fiolhais, Phys. Rev. B
{\bf 46}, 6671 (1992).

\bibitem{mao01} H. K. Mao, J. Xu, V. V. Struzhkin, J. Shu,
R. J. Hemley, L. Vo\v{c}adlo, D. Alf\`e, G. D. Price, M. J. Gillan,
W. Sturham, M. Y. Hu, E. E. Alp, M. Schwoerer-B\"ohning, D,
H\"ausermann, P. Eng, G. Shen, H. Giefers, R. L\"ubbers,
and G. W\"ortmann, Science {\bf 292}, 914 (2001).

\bibitem{vanderbilt90} D. Vanderbilt, Phys. Rev. B {\bf 41},
7892 (1990).

\bibitem{kresse96}
G. Kresse and J. Furthm\"{u}ller, Phys. Rev. B {\bf 54}, 11169 (1996);
G. Kresse and J. Furthm\"{u}ller, Comput. Mater. Sci. {\bf 6}, 15 (1996);
a discussion of the ultra-soft pseudopotentials used in the VASP
code is given in G. Kresse and J. Hafner, J. Phys. Condens. Matter
{\bf 6}, 8245 (1994).

\bibitem{alfe99b}
D. Alf\`{e}, Computer Phys. Commun. {\bf 118}, 31 (1999).

\bibitem{anderson97}
O. L. Anderson and A. Duba, J. Geophys. Res. {\bf 102}, 22659 (1997).

\bibitem{vocadlo00}
L. Vo\v{c}adlo, D. Alf\`{e}, J. P. Brodholt, G. D. Price,
and M. J. Gillan, Phys. Earth Planet.
Inter. {\bf 117}, 123 (2000).

\bibitem{monkhorst76} H. J. Monkhorst and J. D. Pack, Phys. Rev. B
{\bf 13}, 5188 (1976).








\end{thebibliography}
\end{document}